\documentclass[twocolumn,aps, prb, showpacs,preprintnumbers,amsmath,amssymb]{revtex4-1}

\usepackage{graphicx}
\usepackage{dcolumn}
\usepackage{bm}
\usepackage[percent]{overpic}

\begin{document}
\newcommand{\Arg}[1]{\mbox{Arg}\left[#1\right]}
\newcommand{\bb}{\mathbf}
\newcommand{\braopket}[3]{\left \langle #1\right| \hat #2 \left|#3 \right \rangle}
\newcommand{\braket}[2]{\langle #1|#2\rangle}
\newcommand{\be}{\[}
\newcommand{\br}{\vspace{4mm}}
\newcommand{\bra}[1]{\langle #1|}
\newcommand{\braketbraket}[4]{\langle #1|#2\rangle\langle #3|#4\rangle}
\newcommand{\braop}[2]{\langle #1| \hat #2}
\newcommand{\dd}[1]{ \! \! \!  \mbox{d}#1\ }
\newcommand{\DD}[2]{\frac{\! \! \! \mbox d}{\mbox d #1}#2}
\renewcommand{\det}[1]{\mbox{det}\left(#1\right)}
\newcommand{\ee}{\]} 
\newcommand{\eg}{\textbf{\\  Example: \ \ \ }}
\newcommand{\Imag}[1]{\mbox{Im}\left(#1\right)}
\newcommand{\ket}[1]{|#1\rangle}
\newcommand{\ketbra}[2]{|#1\rangle \langle #2|}
\newcommand{\kp}{\arccos(\frac{\omega - \epsilon}{2t})}
\newcommand{\ldos}{\mbox{L.D.O.S.}}
\renewcommand{\log}[1]{\mbox{log}\left(#1\right)}
\newcommand{\Log}{\mbox{log}}
\newcommand{\Modsq}[1]{\left| #1\right|^2}
\newcommand{\nb}{\textbf{Note: \ \ \ }}
\newcommand{\op}[1]{\hat {#1}}
\newcommand{\opket}[2]{\hat #1 | #2 \rangle}
\newcommand{\occ}{\mbox{Occ. Num.}}
\newcommand{\Real}[1]{\mbox{Re}\left(#1\right)}
\newcommand{\so}{\Rightarrow}
\newcommand{\sol}{\textbf{Solution: \ \ \ }}
\newcommand{\thetafn}[1]{\  \! \theta \left(#1\right)}
\newcommand{\tin}{\int_{-\infty}^{+\infty}\! \! \!\!\!\!\!}
\newcommand{\Tr}[1]{\mbox{Tr}\left(#1\right)}
\newcommand{\kb}{k_B}
\newcommand{\rad}{\mbox{ rad}}
\preprint{APS/123-QED}

\title{RKKY interaction between adsorbed magnetic impurities in graphene: \\symmetry and strain effects}

\author{P. D. Gorman$^{(a)}$, J. M. Duffy$^{(a)}$, M. S. Ferreira$^{(a, b)}$ and S. R. Power$^{(c)}$}\email{spow@nanotech.dtu.dk}

\affiliation{a) School of Physics, Trinity College Dublin, Dublin 2, Ireland \\
b) CRANN, Trinity College Dublin, Dublin 2, Ireland \\
c) Center for Nanostructured Graphene (CNG), DTU Nanotech, Department of Micro- and Nanotechnology,
Technical University of Denmark, DK-2800 Kongens Lyngby, Denmark }

\date{\today}

\begin{abstract}
The growing interest in carbon-based spintronics has stimulated a number of recent theoretical studies on the RKKY interaction in graphene, with the aim of determining the most energetically favourable alignments between embedded magnetic moments.
The RKKY interaction in undoped graphene decays faster than expected for conventional two-dimensional materials and recent studies suggest that the adsorption configurations favoured by many transition-metal impurities may lead to even shorter ranged decays and possible sign-changing oscillations.
Here we show that these features emerge in a mathematically transparent manner when the symmetry of the configurations is included in the calculation.
Furthermore, we show that by breaking the symmetry of the graphene lattice, via uniaxial strain, the decay rate, and hence the range, of the RKKY interaction can be significantly altered.
Our results suggest that magnetic interactions between adsorbed impurities in graphene can be manipulated by careful strain engineering of such systems.
\end{abstract}

\pacs{}
                 
\maketitle

\section{Introduction}
\label{intro}
Graphene has been attracting the interest of the wider scientific community due to its potential for applications in fields as diverse as photonics, sensor technology, and spintronics.\cite{riseofgraphene, neto:graphrmp, yazyev:review}
Spintronics is a particularly promising field for graphene application due to the weak spin-orbit and hyperfine interactions, which in other materials act as significant sources of spin relaxation and decoherence.\cite{kane_quantum_2005, huertas-hernando_spin-orbit_2006, min_intrinsic_2006, huertas-hernando_spin-orbit-mediated_2009, trauzettel_spin_2007, yazyev_hyperfine_2008, fischer_hyperfine_2009} 

One recurrent topic in the field of spintronics is the mechanism of interaction between localized magnetic moments embedded in nanoscale systems.
An indirect exchange interaction mediated by the conduction electrons of a host medium manifests as an energy difference between different alignments of the localized moments. 
Such an interaction is usually calculated within the Ruderman-Kittel-Kasuya-Yosida (RKKY) approximation\cite{RKKY:RK, RKKY:K, RKKY:Y}, and the interaction itself frequently takes this name.\cite{RKKYIEC}

The RKKY interaction in graphene has been intensively studied. \cite{stephenreview, Vozmediano:2005, dugaev:rkkygraphene, saremi:graphenerkky, brey:graphenerkky, hwang:rkkygraphene, bunder:rkkygraphene, black:graphenerkky, sherafati:graphenerkky, uchoa:rkkygraphene, black-schaffer_importance_2010, me:grapheneGF, sherafati:rkkygraphene2, kogan:rkkygraphene, disorderedRKKY,klinovaja_rkky_2013}
The consensus from these studies is that the interaction strength decays asymptotically as $D^{-3}$ in undoped graphene, where $D$ is the separation between magnetic moments.
This decay rate is faster than the $D^{-2}$ decay expected for conventional two-dimensional materials and arises from the vanishing density of states at the Fermi energy in graphene.\cite{me:grapheneGF}
The usual sign-changing oscillations predicted for such interactions are masked by the coincidence of the Fermi surface and Brillouin zone.
This causes the sign of the coupling - which determines the ferromagnetic (FM) or antiferromagnetic (AFM) alignment of the moments - within the RKKY interaction to only depend on whether the two moments occupy the same or opposite sublattices, and not on their separation.
When graphene is doped or gated such that the Fermi surface no longer coincides with the Brillouin zone, sign-changing oscillations are recovered and the interaction is found to decay as $D^{-2}$. 

\begin{figure}
\includegraphics[width=0.45\textwidth]{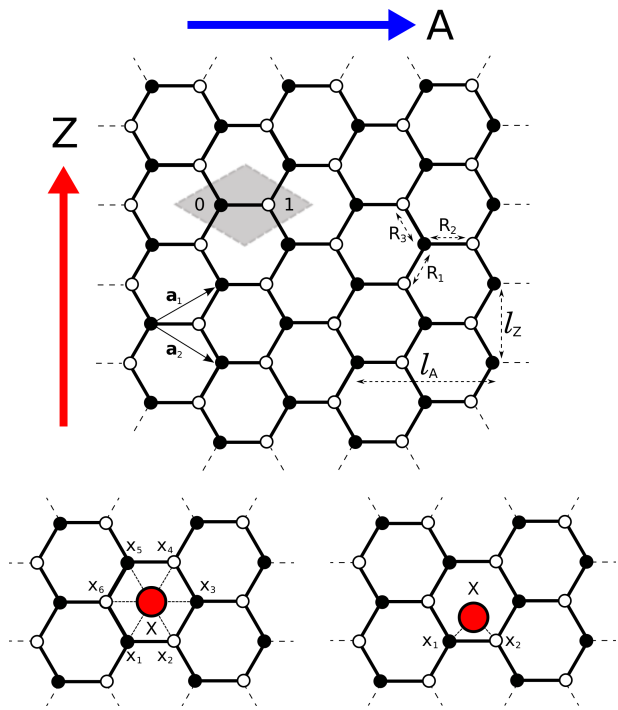}
\caption{
Schematic representation of the graphene lattice showing with the armchair ($A$) and zigzag ($Z$) directions and units of separation ($l_A$ and $l_Z$), the two-atom unit cell (shaded area) and lattice vectors $\mathbf{a}_1$ and $\mathbf{a}_2$, and the bond lengths $R_1$, $R_2$ and $R_3$ between an atom on the lattice and its nearest neighbours.
The filled and hollow symbols represent sites on different sublattices. The bottom panels show magnetic impurities ($X$) attached to graphene lattice atoms $x_i$ in the center-adsorbed (left) and bridge-adsorbed (right) configurations.
}
\label{fig:schematic}
\end{figure}

Some studies have extended the discussion to include center-adsorbed impurities and bridge-adsorbed impurities (Fig. \ref{fig:schematic}).
Center-adsorbed impurities (often called plaquette or `hollow-site' impurities) consist of an impurity atom located at the center of a hexagon in the graphene lattice, connected symmetrically to the six surrounding carbon atoms.
Bridge-adsorbed impurities (often called bond impurities) consist of an impurity atom located above the bond between two adjacent carbon atoms in the graphene lattice, connected symmetrically to both.
These types of adsorption are of particular interest since they are energetically favourable for many transition-metal atoms, with the majority preferring the center-adsorbed configuration.\cite{eelbo_adatoms_2013, kengo_nakada_akira_ishii_dft_2011,lehtinen_magnetic_2003}
There is some discrepancy in the literature about the basic features of the interaction between center-adsorbed impurities.
Some studies suggest an interaction which is always AFM and decays as $D^{-3}$ (the same decay rate predicted for substitutional) while others suggest a decay rate of $D^{-7}$ with a FM interaction at some separations.\cite{saremi:graphenerkky, sherafati:graphenerkky,uchoa:rkkygraphene}
This discrepancy is similar to the related case of carbon nanotubes, where center-adsorbed impurities are predicted to display a decay rate of $D^{-5}$ instead of $D^{-1}$ found for substitutional impurities. \cite{AntonioDavidIEC, David:IEC, DavidSpinValve}

Recent interest in the strain engineering of graphene is motivated by the high degree of tunability that can be achieved by varying the strength and type of mechanical strain applied.\cite{Pereira09, pereira_tight-binding_2009, Guinea:gapsgraphene,sharma_effect_2013}
The ability of graphene to sustain reversible deformations of up to approximately 20\% \cite{PhysRevB.76.064120} suggests that even simple uniaxial strains may provide opportunities to tune the electronic, magnetic, optical and thermal properties of graphene systems.
Strain has recently been predicted to significantly modify features of the interaction between substitutional magnetic moments embedded in graphene.\cite{ourpaper,Peng20123434}
Due to the number of transition-metal atoms that adsorb in either the center or bridge configurations, and the different features observed for these configurations, it is worth expanding this previous work to predict the effects of strain on the interactions between adsorbed impurities.
Since a range of effects are predicated on exchange interactions, the ability to manipulate these interactions via strain may lead to interesting spintronic-applications.

The remainder of this paper is organised as follows.
In section \ref{RKKY} we introduce the theoretical prescription to describe the graphene-impurity system, calculate the RKKY interaction in terms of single-particle Green functions (GFs), and provide an analytical approach to extract the decay behaviour using the Stationary Phase Approximation (SPA) and RKKY approximation.
In section \ref{sec:results} we compare numerical calculations and analytical predictions of the energy difference between FM and AFM alignments of the moments.
In section \ref{strain} we explore how breaking the symmetry of the system via uniaxial strain leads to longer ranged interactions for center-adsorbed impurities, and strain-controlled sign changes in the coupling of bridge-adsorbed impurities.
And in section \ref{conclude} we discuss our results and their implications.

\section{Methods}
\label{RKKY}
The indirect exchange coupling between two moments embedded in a conducting host can be calculated by considering the energy difference between the ferromagnetic (FM) and antiferromagnetic (AFM) alignments of the moments.\cite{Lloyd,stephenreview}
The total energy difference, $J_{BA}$, between two magnetic impurities labeled $A$ and $B$, can be calculated using the Lloyd formula method
\begin{equation}
J_{BA} = -\frac{1}{\pi} \,\mathrm{Im}  \int  \mathrm{d} E \, f(E) \, \ln  \left( 1 + 4 \, V_{ex}^2 \,  \mathcal{G}_{BA}^{\uparrow} (E) \,  \mathcal{G}_{AB}^{\downarrow} (E) \right) \,,
\label{staticJ}
\end{equation}
where $ \mathcal{G}_{AB}^{\sigma} (E)$ is the real-space, single-electron Green Function (GF) describing the propagation of electrons with spin $\sigma = \uparrow$ or $\downarrow$, $V_{ex}$ is the exchange splitting of the magnetic impurity and $f(E)$ is the Fermi function.

To calculate the required GFs we employ an Anderson-like Hamiltonian\cite{anderson_localized_1961} to describe the electronic properties of the system, whose general form is given by
\begin{equation}
\begin{split}
 \hat{H} &=   \sum_{\langle j, \ell \rangle, \sigma} t_{j, \ell} \, \ {\hat c}_{j \sigma}^\dag \, {\hat c}_{\ell \sigma} \\ & + \sum_{X, \sigma} \left( \epsilon_X^\sigma  \ {\hat c}_{X\sigma}^\dag \, {\hat c}_{X\sigma} + \sum_{x = x_1 }^{x_N} \left( \tau_{X̣, x}  {\hat c}_{X \sigma}^\dag \, {\hat c}_{x \sigma} + c.c. \right)  \right) \,.
 \label{hamiltonian}
 \end{split}
\end{equation}
Here ${\hat c}_{j \sigma}^\dag$ (${\hat c}_{j \sigma}$) creates (annihilates) an electron with spin $\sigma$ in a $\pi$ orbital centerd at site $j$ in the graphene lattice, $t_{j, \ell}$ is the electronic hopping term between two such orbitals, where $t_{j, \ell}=0$ if sites $j$ and $l$ are not nearest neighbours. 
The first term in Eq. \eqref{hamiltonian} is thus simply the nearest-neighbour tight-binding (NNTB) Hamiltonian for the pristine graphene lattice, with nearest neighbour hopping $t = -2.7 eV$. 
The second term provides a simple description of the magnetic impurity orbitals ($X = A, B$) and their connection to the lattice.
We assume that each impurity orbital has a finite hopping, $\tau_{X, x}$, to $N$ of the carbon $\pi$ orbitals located at sites $x = \{x_1, \cdots, x_N\}$ surrounding the impurity. 
The specific cases we consider in this paper are $N=6$ (center-adsorbed) and $N=2$ (bridge-adsorbed), as shown in the bottom panel of Fig. \ref{fig:schematic}.
The quantity $\epsilon_X^\sigma = \mp V_{ex}$ is a spin-dependent onsite potential that accounts for the exchange splitting in the magnetic orbitals. 
In this model, we consider only a single magnetic orbital at each impurity site. 
However, it is straightforward to generalise the approach to deal with multiple orbitals.
The exact parametrizations for specific impurity types can be found by comparison to \emph{ab initio} studies of single impurities adsorbed onto a graphene sheet, which have been performed for a wide range of impurity species with different adsorption configurations.\cite{kengo_nakada_akira_ishii_dft_2011, yagi_stable_2004, sevincli_electronic_2008, cao_transition_2010, chan_first-principles_2008, gao_first-principles_2010, eelbo_adatoms_2013, duffy_magnetism_1998, mao_density_2008, lehtinen_magnetic_2003}

\subsection{Green functions}
The Green function matrix elements, $ \mathcal{G}_{AB}^{\sigma} (E)$, required for the calculation in \mbox{Eq. \eqref{staticJ}} are obtained using the Dyson equation.
This allows the complete Green function to be written in terms of the pristine lattice GFs of the graphene lattice ($g_{ab}$) associated with the first term of the Hamiltonian in \mbox{Eq. \eqref{hamiltonian}}.
We find
\begin{equation}
 \mathcal{G}_{AB} = \frac{g_{AA} \, \Gamma_{AB} \,  g_{BB}}{(1 - g_{AA} \Gamma_{AA})(1 - g_{BB} \Gamma_{BB}) - g_{AA} \Gamma_{AB} g_{BB} \Gamma_{BA} },
 \label{eq:GAB}
\end{equation}
where $g_{AA}$ is the GF for the disconnected impurity and we define $\Gamma_{AB}$ to be the sum of the $N^2$ pristine graphene GF matrix elements connecting the two impurity sites
\begin{equation}
 \Gamma_{AB} = \sum_{a=a_1}^{a_N} \sum_{b=b_1}^{b_N}  \tau_{Aa} \, g_{ab} \, \tau_{bB} \equiv \tau^2 \, \sum_{a,b} \, g_{ab} \,.
\label{eq:gamma}
\end{equation}
In the last term of Eq. \eqref{eq:gamma} we assume that the hopping terms from each of the connecting sites to the impurity orbital are identical.
The on-site potentials required for the spin-dependent GFs can be added similarly using the Dyson equation.

We note that the only term in Eq. \eqref{eq:GAB} that depends on the separation between $A$ and $B$ is $\Gamma_{AB}$. 
Since the second term in the denominator of Eq. \eqref{eq:GAB} decays rapidly for appreciable separations, it is clear that $\mathcal{G}_{AB} ( D )\sim\Gamma_{AB} ( D )$.
Thus we expect $\Gamma_{AB}$ to dominate in our investigation of the coupling and we now examine the form of this quantity.
The pristine GFs, $g_{ab}$, appearing in Eq. \eqref{eq:gamma} can be calculated using the Bloch theorem to avail of the periodicity of the pristine graphene lattice.
The GF connecting two sites on the graphene lattice in unit cells separated by a vector $\mathbf{D}$ can be written as a double integral over the Brillouin Zone in reciprocal space 
\begin{equation}
 g_{ab} = \frac{1}{2\pi^2} \int\limits_{-\pi/2}^{\pi/2} dk_Z \int\limits_{-\pi}^{\pi} dk_A \, \frac{N_{ab}(E,\mathbf{k}) \, e^{i \mathbf{k} \cdot \mathbf{D}}}{E^2 -t^2 \, |f(\mathbf{k})|^2},
 \label{eq:gab_int}
\end{equation}
where $k_A = \tfrac{\sqrt{3} k_x a}{2}$ and $k_Z = \tfrac{k_y a}{2}$ are dimensionless wavevectors in the armchair and zigzag directions respectively and 
\begin{equation}
 f(\mathbf{k}) = 1 + 2 \cos(k_Z) e^{i k_A}
 \label{eq:fk}
\end{equation}
is related to the electronic dispersion relation of the NNTB Hamiltonian by $\epsilon_\pm = \pm t | f (\mathbf{k}) |$.
$N_{ab}(E, \mathbf{k})$ is a complex function whose exact form depends on whether the sites $a$ and $b$ belong to the same or opposite sublattices (represented schematically by filled and hollow circles in Fig. \ref{fig:schematic}) and is given by
\begin{equation}
 N_{ab}(E, \mathbf{k}) = \begin{cases}       	
	E		  &\mbox{for }  \{a, b\} \in  \{\bullet,\bullet\} \vee \{\circ, \circ\} \\
	tf(\mathbf{k})	    	  & \mbox{for }  \{a, b\} \in  \{\bullet, \circ\}  \\
	tf^{\star}(\mathbf{k})	  & \mbox{for }  \{a, b\} \in  \{\circ, \bullet\}  
\end{cases}.
\end{equation}
The numerical cost of evaluating graphene GFs using Eq. \eqref{eq:gab_int} can be reduced considerably by noting that either of the two integrals can first be performed analytically using contour integration.\cite{me:grapheneGF} 

From Eq. \eqref{eq:gamma} it is clear that $\Gamma_{AB}$ can be written as a sum of $N^2$ integrals.
However, for numerical and analytical convenience it is useful to take the summation inside the integrals before they are solved.
We can then write
\begin{equation}
 \Gamma_{AB}	= \frac{1}{2\pi^2} \int\limits_{-\pi/2}^{\pi/2} dk_Z \int\limits_{-\pi}^{\pi} dk_A \frac{ \mathcal{M} (E,\mathbf{k})\, e^{i \mathbf{k} \cdot \mathbf{D}} }{E^2 -t^2 |f(\mathbf{k})|^2} \,,
 \label{GammaAB_M}
\end{equation}
where $\mathbf{D}$ here is the separation vector between the impurities $A$ and $B$, or indeed, between any two equivalent sites $a_i$ and $b_i$ around each impurity site.
$\mathcal{M} (E, \mathbf{k})$ takes into account the net effect of the various $N_{ab}$ and additional phase terms that arise during the summation over $a$ and $b$ and is given by 
\begin{equation}
 \mathcal{M}(E,\mathbf{k})=  \sum_{a=a_1}^{a_N} \sum_{b=b_1}^{b_N} N_{ab}(E,\mathbf{k}) \, e^{i \, \mathbf{k}\cdot\,(\mathbf{D}_{ba} - \mathbf{D}) } \,,
\label{eq:mek}
\end{equation}
where $\mathbf{D}_{ba}$ is the separation vector between the unit cell containing the site $a$ connecting to impurity $A$ and that containing site $b$ connecting to $B$.
The form of $\mathcal{M}(E,\mathbf{k})$ thus depends on the nature of the impurity and its connection to the graphene lattice.
Explicit expressions for the center- and bridge-adsorbed cases will be given in Section \ref{sec:results}. 
We note that \mbox{Eq. \eqref{eq:gab_int}} for $g_{ab}$ and Eq. \eqref{GammaAB_M} for $\Gamma_{AB}$ are very similar in form, with $\mathcal{M}(E,\mathbf{k})$ taking the place of $N_{ab}$ in the latter.
It is thus instructive to examine whether methods that have proven useful for the single-site GFs can also be availed of when the multi-site $\Gamma_{AB}$ term is of interest.
Firstly, we note that once more contour integration can be used to perform one of the two integrals in \mbox{Eq. \eqref{GammaAB_M}}.
Numerical tests confirm that identical results are obtained whether $\Gamma_{AB}$ is evaluated using the single or double numerical integration methods or using a summation of the single site GFs given in \mbox{Eq. \eqref{eq:gamma}}.
The methods we have introduced thus far have reduced the calculation of $\Gamma_{AB}$ for center-adsorbed impurities from a sum of $36$ two-dimensional integrals to just a single one-dimensional integral, allowing much faster numerical evaluation of $\Gamma_{AB}$ and quantities, such as the magnetic coupling, which rely upon it.
We have shown previously that the pristine graphene GFs between sites separated along the high-symmetry directions are very well approximated throughout the entire energy band using the Stationary Phase Approximation (SPA).\cite{ourpaper}
This method takes advantage of the highly oscillatory nature of the integrand and approximates the integral near stationary points, $k^0$, where the oscillations are slowest.
It returns a closed-form analytic expression for the GF, which we have previously applied to studies of both the standard RKKY interaction\cite{me:grapheneGF} and dynamic spin excitations of substitutional magnetic impurities in graphene.\cite{DynamicRKKY}
Using the SPA approach, the off-diagonal element of the graphene lattice GF between two sites on the same sublattice can be written as a sum of terms of the form
\begin{equation}
 g_{ab}(E) = \frac{\mathcal{A}(E)e^{i \mathcal{Q}(E)D}}{\sqrt{D}}\,,
 \label{g_SPA}
\end{equation}
where $\mathcal{A}(E)$ is an energy-dependent coefficient and $\mathcal{Q}(E)$ can be identified with the Fermi wave vector in the direction of separation.
The exact functional forms of these quantities depend on the separation direction, but the distance dependence is clear in this form. An analogous expression can be derived for $\Gamma_{AB}$.
Since the oscillatory terms in the integrands for $g_{ab}$ and $\Gamma_{AB}$ are identical, the stationary points occur at exactly the same values.
Thus, the only alteration made to Eq. \eqref{g_SPA} is to the coefficient $\mathcal{A}(E)$, which is multiplied by a factor $\frac{\mathcal{M}_0 (E)}{E}$, where $\mathcal{M}_0$ is found by evaluating Eq. \eqref{eq:mek} at the stationary point.
Explicit expressions for the stationary points and for the coefficients $\mathcal{A}(E)$ and $\mathcal{Q}(E)$ are calculated in Ref. [\onlinecite{me:grapheneGF}] for the high symmetry armchair and zigzag directions, and will be used in later sections to calculate the analytic form of $\Gamma_{AB}$ for center-adsorbed and bridge-adsorbed impurities with these separation directions.

\subsection{RKKY interaction}
\label{section_methods_rkky}
Numerical calculations of the indirect exchange coupling within this work are performed by evaluating the integral in \mbox{Eq. \eqref{staticJ}} with the full Green functions calculated using \mbox{Eqs. \eqref{eq:GAB} - \eqref{eq:mek}}.
To explore the behaviour of the interaction analytically, it is worth noting that for small exchange splittings $V_{ex}$, the logarithm in \mbox{Eq. \eqref{staticJ}} can be approximated by the leading term in a Taylor expansion so that the coupling becomes
\begin{equation}
 J_{BA} \approx - \frac{4 \; V_{ex}^2 }{\pi} \:\mathrm{Im} \: \int \: \mathrm{d} E \: f(E) \,  \mathcal{G}_{AB}^2 (E)  \,.
\end{equation}
This expression is equivalent to the commonly used RKKY approximation, where we note that the spin-dependent GFs are replaced by their spin-independent counterparts.
For substitutional impurities, the pristine graphene lattice GFs are used and the expression is rewritten in terms of the spin susceptibility, $\chi$.
For adsorbed atoms, we have seen that the separation dependent behaviour of the full GF is determined by that of $\Gamma_{AB}$ and so we make the additional approximation
\begin{equation}
 J_{BA} \sim - {V_{ex}^2 }\:\mathrm{Im} \: \int \: \mathrm{d} E \: f(E) \,  {\Gamma}_{AB}^2 (E)  \,,
 \label{eq:rkkysum}
\end{equation}
which encapsulates all the relevant separation-dependent behaviour of the interaction between adsorbed impurities.
Within the SPA approach, we have seen above that ${\Gamma}_{AB}$ can be written in a form analogous to Eq. \eqref{g_SPA}
\begin{equation}
 \Gamma_{AB} (E) = \frac{\mathcal{A}_{\Gamma}(E)e^{i \mathcal{Q}(E)D}}{\sqrt{D}},
 \label{Gamma_SPA}
\end{equation}
where $\mathcal{A}_{\Gamma}(E)$ is related to $\mathcal{A}(E)$ in \mbox{Eq. \eqref{g_SPA}}.

We have also shown previously that the behaviour of the magnetic coupling can be extracted quite easily when the GFs are expressed in such a form.
The integration procedure can be reduced to a sum over Matsubara frequencies and when the functions $\mathcal{B} (E) = \mathcal{A}_{\Gamma}^2(E)$ and $\mathcal{Q}(E)$ are expanded around the Fermi energy in the low temperature limit we find 
\begin{equation}
 J_{BA} \sim \text{Im} \sum_{\ell=0} \frac{\mathcal{J}_\ell(E_F)}{D^{\ell+2}}e^{i 2 \mathcal{Q}(E_F)D},
\end{equation}
where
\begin{equation}
\label{eq:Jl}
 \mathcal{J}_\ell(E_F) = \frac{V_{ex}^2 \mathcal{B}^{(\ell)}(E_F)}{[2 i Q^{(1)}(E_F)]^{\ell+1}}
\end{equation}
is the distance-independent coefficient for the $\ell^{th}$ term in the series, $\ell$ is a non-negative integer and $\mathcal{B}^{(\ell)}$ is the $\ell^{th}$ order energy derivative of $\mathcal{B}(E)$ evaluated at $E_F$.
From this definition it should be clear that the leading term in this series (the first non-zero $\mathcal{B}^{(\ell)}$) determines the asymptotic decay rate of the coupling, which goes as $1/D^{\ell+2}$.
For substitutional impurities in graphene it is found that the $\ell = 0$ term vanishes, leading to a decay rate of $J \sim D^{-3}$, faster than expected for a two-dimensional material. 

In the following sections, we will show the explicit form of the expressions derived above for the specific cases of center-adsorbed and bridge-adsorbed impurities.
We examine some of the features of $\Gamma_{AB}$ in each case and show how they lead to interesting results for the interactions between magnetic impurities which adsorb in these configurations.

\section{Impurity configurations}
\label{sec:results}
\subsection{Center-adsorbed impurities}
\label{plaquette}
\begin{figure}
\includegraphics[width=0.45\textwidth]{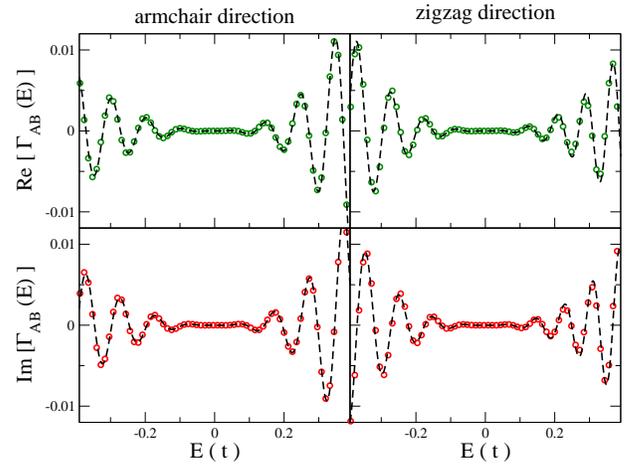}
\caption{
Numerical (symbols) and analytical (dashed lines) evaluations of the real (top panels) and imaginary (bottom panels) components of $\Gamma_{AB}$ for two center-adsorbed type impurities with separations of $30\ l_A$ in the left panels and $60\ l_Z$ in the right panels.
}
\label{fig:plaquette_gamma}
\end{figure}

Center-adsorbed impurities are of particular interest in the study of magnetically-doped graphene since this configuration is the most energetically favourable for the majority of single-atom impurities, including many transition-metal atoms such as Fe, Mn and Co. \cite{kengo_nakada_akira_ishii_dft_2011, eelbo_adatoms_2013, mao_density_2008, duffy_magnetism_1998, cao_transition_2010, sevincli_electronic_2008, chan_first-principles_2008}
Each center-adsorbed impurity is connected to the 6 surrounding atoms in the lattice, as shown in the bottom left panel of \mbox{Fig. \ref{fig:schematic}}, so that the sum in \mbox{Eq. \eqref{eq:mek}} consists of 36 terms.
The symmetry of many of these terms allows much simplification and we can write 
\begin{equation}
 \mathcal{M}_C(E,\mathbf{k}) = 2 E |f(\mathbf{k})|^2 +  2 t \, \text{Re} \,\left[{ f^3(\mathbf{k}) \, e^{-i2k_A}}\right].
 \label{eq:mplaq}
\end{equation}
Using this expression in conjunction with \mbox{Eq. \eqref{GammaAB_M}} provides an efficient method to calculate $\Gamma_{AB}$ numerically for center-adsorbed impurities, especially when contour integration is used to reduce the numerical evaluation to a one-dimensional integral in reciprocal space.
When using the contour integration approach, the correct sign of the pole must be taken in each term of $\mathcal{M}(E,\mathbf{k})$ and it is usually necessary to split up some of the trigonometrical expressions into their exponential components to achieve an exact match with the sum of individual GFs.
To gain an insight into the analytic behaviour of $\Gamma_{AB}$ for large separations, we can loosen these constraints and evaluate the $\mathcal{M}_C(E,\mathbf{k})$ term within the SPA approximation in the high symmetry armchair and zigzag directions.
For armchair separations, a single stationary point is sufficient for a very accurate approximation in the energy range $|E| < |t|$. At this stationary point we find 
\begin{equation}
 \mathcal{M}_{C}^{ac}(E,k^0) =\frac{2E^3 (t -E) }{t^3} .
\end{equation}
Generalising the single-site SPA GF derived in Ref. \onlinecite{me:grapheneGF}, we find the following coefficients for \mbox{Eq. \eqref{Gamma_SPA}}
\begin{equation}
\begin{split}
 \mathcal{A}_{C}^{ac}(E) &= \tau^2\,\sqrt{\frac{2 i }{\pi}} \, \sqrt{ \frac{E}{(E^2 + 3t^2)\sqrt{(t^2 - E^2)}}}\, \frac{2E^2 (t -E) }{t^3}\\
 \mathcal{Q}^{ac}(E) &= \pm \cos^{-1}\left({ \frac{-\sqrt{t^2 - E^2}}{t}  }\right),  \\ 
\end{split}\,
\label{eq:armchair_aq}
\end{equation}
where we note that the value of $\mathcal{Q}$ is identical to the single impurity case so we omit the $C$ subscript.
The choice of sign for $\mathcal{Q}^{ac}$ emerges from the requirement that the poles involved in the contour integration lie within the chosen contour, and for positive separations in the armchair direction it is the sign that obeys the constraint $\text{Im}\,[{Q}^{ac}(E)] > 0$.
Eqs. \eqref{eq:armchair_aq} and \eqref{Gamma_SPA} provide a closed form analytical expression for $\Gamma_{AB}$ for armchair separated center-adsorbed impurities.
The left-hand side panels of Fig. \ref{fig:plaquette_gamma} show a comparison of this quantity with a complete numerical evaluation for a separation of $30\ l_A$ and we note an excellent agreement, confirming the validity of the SPA approach. 

A similar approach can be followed for zigzag separations, again following the prescription given in Ref \onlinecite{me:grapheneGF}.
We note that, for this direction, there are generally two contributing terms of the type shown in \mbox{Eq. \eqref{g_SPA}} which must be considered when deriving the SPA GF.
Each has a corresponding evaluation for $\mathcal{M}_{C}^{zz}(E,k^0)$. However, one of these evaluations is identically zero, such that only one of the contributions needs to be considered.
The surviving value is
\begin{equation}
 \mathcal{M}_{C}^{zz}(E,k^0) =\frac{4E^3}{t^2} \,,
\end{equation}
and the corresponding SPA coefficients are
\begin{equation}
\begin{split}
 \mathcal{A}_C^{zz}(E) &= \frac{\tau^2}{\sqrt{2 i \pi}}\frac{4E^2}{t^2} \sqrt{ \frac{E}{|t|(t - E)\sqrt{(4t^2 -(E - t)^2)}}} \\
 \mathcal{Q}_{C}^{zz}(E) &= \pm \cos^{-1}\left({ \frac{-t + E}{2t}  }\right),\\
\end{split}\,
\label{eq:zigzag_aq}
\end{equation}
where $\mathcal{Q}_{C}^{zz}(E)$ is the same as for one of the single-impurity zigzag-direction cases and again has a sign choice emerging from the contour integration. 

\begin{figure}
\includegraphics[width=0.48\textwidth]{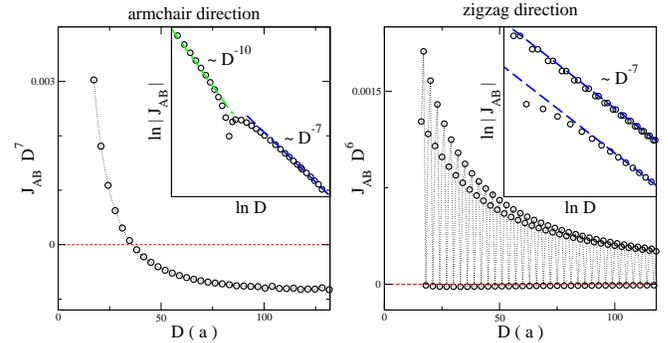}
\caption{
Numerical evaluation of the coupling between center-adsorbed impurities as a function of separation, $D$, in the armchair (left) and zigzag (right) directions.
The armchair (zigzag) results are multiplied by $D^7$ ($D^6$) to highlight the features discussed in the text. The red dashed line in the main panels highlights the boundary between AFM (above) and FM (below) couplings.
In the armchair direction an initial AFM interaction decays extremely rapidly as $D^{-10}$ before a sign change to FM and a decay of $D^{-7}$ at larger separations.
Zigzag separations reveal that every third value of separation has an FM interaction approximately 2 orders of magnitude smaller than the AFM majority values.
The insets in each case show log-log plots where dashed lines show the slopes corresponding to the relevant decay rates. 
}
\label{fig:plaquette_coupling}
\end{figure}

From the SPA coefficients we can predict the decay rates for the RKKY interaction between two center-adsorbed magnetic impurities.
To determine the decay rate we must determine the first non-vanishing energy derivative, $\mathcal{B}^{(\ell)}$ of $\mathcal{B} = \mathcal{A}^2$, evaluated at the Fermi energy $E_F = 0.0$.
Using the expressions for $\mathcal{A}$ in Eqs. \eqref{eq:armchair_aq} and \eqref{eq:zigzag_aq}, this is found to occur at $\ell = 5$, corresponding to a decay rate of $J \sim D^{-7}$, for both armchair and zigzag separations of center-adsorbed impurities.
This is significantly faster than the $J \sim D^{-3}$ rate predicted for substitutional impurities in graphene, or the more general $J \sim D^{-2}$ rate predicted for two dimensional materials.
This point will be discussed in further detail in Sec. \ref{strain-plaq-sec}, when strain is introduced.
Comparing these predictions with numerical calculations of the complete exchange interaction reveals a more complicated picture (Fig. \ref{fig:plaquette_coupling}).
The first point to note is that a much faster decay rate than the substitutional case is noted for all directions, and in the zigzag direction a decay of $D^{-7}$ is noted in agreement with the analytic prediction.
However in the armchair direction, an even faster decay of approximately $D^{-10}$ is noted initially leading to a sign change, with a decay of $D^{-7}$ recovered in the asymptotic limit.
Thus our analytic result captures the large separation limit in each direction.
An interesting point to note is also that the sign of the interaction is not AFM at all sites, as has been previously predicted for this type of impurity.\cite{saremi:graphenerkky, black:graphenerkky, sherafati:graphenerkky}
In the zigzag direction we note that every third value of separation corresponds to a preferential FM coupling, but that this coupling is approximately two orders of magnitude smaller than the AFM values at similar distances.
The period-3 behaviour for zigzag direction separations is a common feature in graphene and arises due to the form of the component of the Fermi wavevector in this direction.
In the armchair direction, this period-3 behaviour does not arise and a smoother curve is found.
The interaction is initially antiferromagnetic where it decays even more rapidly than predicted, before a sign change gives a very weak ferromagnetic interaction with a $D^{-7}$ decay rate in the asymptotic limit.
For directions between the high symmetry armchair and zigzag directions, a combination of these features is reported as each separation consists of an armchair and zigzag component.
Due to the extremely rapid rates of decay, the interaction between center-adsorbed magnetic impurities is essentially zero for any reasonable separation above a few lattice spacings.
A similar increase in the decay rate has been noted previously for center-adsorbed impurities in carbon nanotubes, but the decay rate here is even more rapid.\cite{David:IEC}
This result would appear to have serious negative implications for spintronic devices aiming to exploit RKKY-like interactions between transition-metal adsorbates in graphene.
We note that although our model assumes equal hopping parameters between the magnetic impurity and the six surrounding carbon atoms, it can be easily shown that the fast decay rate will result as long as the hopping terms to sites on the same sublattices are equivalent.
A similar conclusion is reported in Ref. [\onlinecite{uchoa:rkkygraphene}].

\subsection{Bridge-adsorbed impurities}
\label{bond}
We move our attention now to the case of bridge-adsorbed impurities shown schematically in the bottom right panel of Fig. \ref{fig:schematic}, where the magnetic atom is connected to two neighbouring carbon atoms on the graphene lattice - one from each of the sublattices.
A number of transition-metal atoms are known to favour this configuration over the more common center-adsorbed position.\cite{kengo_nakada_akira_ishii_dft_2011, lehtinen_magnetic_2003, yagi_stable_2004, gao_first-principles_2010}
We can divide pairs of bridge impurities into three classes, depending on the relative orientations of the carbon-carbon bonds over which they are positioned.
Without loss of generality, we assume that one of the impurities is connected over the bond connecting two carbon sites in the same unit cell ($R_2$ in Fig. \ref{fig:schematic}).
The class to which a pair of impurities belong then depends on which of the three possible bond orientations, denoted by $R_1$,  $R_2$ and $R_3$ in Fig. \ref{fig:schematic}, the second impurity is positioned over.
For the current discussion, we will focus on the case when the second impurity also connects to two atoms in the same unit cell, i.e. is also positioned over the $R_2$ bond.
However the behaviour of the other two classes is qualitatively similar.
The Green function connecting two such bridge-adsorbed impurities can be calculated analogously to that for center-adsorbed impurities using \mbox{Eqs. \eqref{eq:GAB}}, \eqref{GammaAB_M} and \eqref{eq:mek}, where taking the summations in \mbox{Eq. \eqref{eq:mek}} over the two atoms at each site we find
\begin{equation}
\label{MEk_bond}
  \mathcal{M}_B(E,\mathbf{k}) = 2 E + 2t \, \text{Re} \, \left[f(\mathbf{k})\right] \,.
\end{equation}
We can use this expression, as in the center-adsorbed case, to either make a full numerical evaluation of the Green function more efficient or within the SPA to get an approximate form of the Green function at large separations.
Within the SPA, we find expressions for $\Gamma_{AB}$ of the form given by \mbox{Eq. \eqref{Gamma_SPA}}, where the coefficients for armchair and zigzag separations are given by
\begin{equation}
\label{eq:bond_aqs}
 \begin{split}
   \mathcal{A}_{B}^{ac}(E) &= \tau^2\,\sqrt{\frac{2 i }{\pi}} \, \sqrt{ \frac{E}{(E^2 + 3t^2)\sqrt{(t^2 - E^2)}}}\, \frac{2(E + t) }{t} \\
 \mathcal{Q}^{ac}(E) &= \pm \cos^{-1}\left({ \frac{-\sqrt{t^2 - E^2}}{t}  }\right)  \\ 
 \mathcal{A}_B^{zz}(E) &= \frac{4 \tau^2} {\sqrt{2 i \pi}} \, \sqrt{ \frac{E}{|t|(t - E)\sqrt{(4t^2 -(E - t)^2)}}} \\
 \mathcal{Q}_{B}^{zz}(E) &= \pm \cos^{-1}\left({ \frac{-t + E}{2t}  }\right) \,,
 \end{split}
\end{equation}
where the sign choices once more relate to the contour integration.
\begin{figure}
\includegraphics[width=0.46\textwidth]{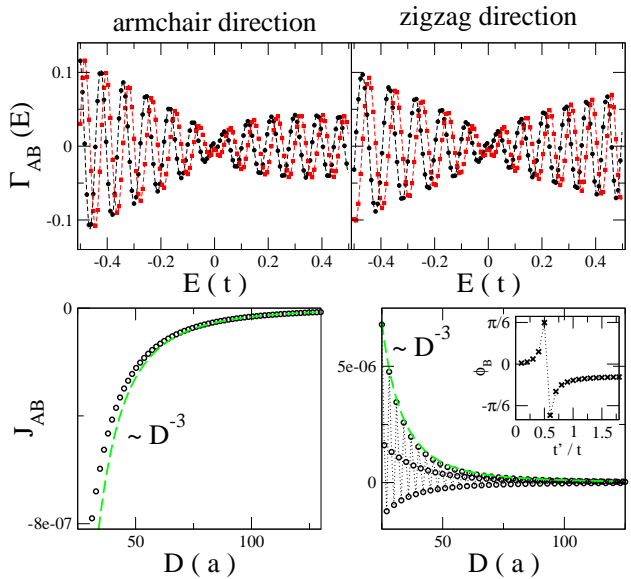}
\caption{
$\Gamma_{AB}$ (top) and coupling (bottom) for bridge-adsorbed impurities separated in the armchair (left) and zigzag (right) directions.
An excellent match is noted between numerical (symbols) and analytic (lines) results for $\Gamma_{AB}$ for separation of $35\ l_A$ and a separation of $60\ l_Z$ for both real (black) and imaginary (red) components.
A monotonically decaying $D^{-3}$ FM interaction is seen in the armchair direction for the class of bridge adsorbates investigated, whereas a sign-changing oscillation is observed in the zigzag case.
The phase of the oscillations is found to vary with the hopping parameter between the impurities and the carbon atoms, as shown in the inset.
}
\label{fig:bridge}
\end{figure}
These expressions are in excellent agreement with numerical evaluations of $\Gamma_{AB}$ for large separations between the bridge-adsorbed impurities, as shown in the top panels of Fig. \ref{fig:bridge} for both high-symmetry directions.
The SPA coefficients also allow us, as before, to predict the decay rate of the RKKY interaction between bridge-adsorbed impurities.
For both directions, the first derivative of $\mathcal{B}$ is non-zero, corresponding to a decay rate of $J_{AB} \sim D^{-3}$, the same rate as predicted for substitutional and top-adsorbed impurities.
The fully numerical calculations shown in the bottom panels of Fig. \ref{fig:bridge} confirm this decay rate but also illustrate additional features.
The armchair case is very similar to the substitutional behaviour, displaying a monotonic $D^{-3}$ decay.
However, it is interesting to note that the interaction in this case is FM.
The other two classes of bridge impurity in this direction (not shown here) have monotonic AFM interactions. 
This is consistent with the interesting behaviour in the zigzag direction, where the usual period-3 oscillation in this direction now displays a sign changing behaviour, with one third of the separations corresponding to preferential FM alignments.
Of the other two classes of pairs of bridge-adsorbed impurities, one displays similar behaviour to that shown here whereas the remaining class shows two-thirds of separations preferring FM alignments.
Thus one-third of the total possible bridge-adsorbed impurity pairs display FM alignments.
In contrast to the center-adsorbed case, the FM interactions have the same order of magnitude as the AFM interactions and the coupling for each class can written as
\begin{equation}
 J_{AB} \sim \frac{1 - 2 \cos (2 \mathcal{Q} D + \phi_B)}{D^3},
 \label{bond_oscills}
\end{equation}
where $\phi_B$ is a phase factor.
This is in contrast to the substitutional case where a non-sign changing oscillation $1 + 2 \cos (2 \mathcal{Q} D )$ is found.
The oscillatory form of the bridge-adsorbed impurity coupling in \mbox{Eq. \eqref{bond_oscills}} has been calculated within the RKKY approximation previously in Ref. [\onlinecite{sherafati:graphenerkky}].
Here it is associated with a direction-dependent phase factor that arises between the interactions when the moments are on the same or on opposite sublattices.
An interesting feature is that the phase of the oscillation between bridge-adsorbed impurities, $\phi_B$, depends on the hopping parameter, $\tau$, connecting the impurity to the two neighbouring carbon atoms.
The form of this dependence is shown in the inset of Fig. \ref{fig:bridge}.
This means that different impurity species will have different oscillation phases and may make feature detection difficult when only a small number of separation values are available, for example in DFT calculations.

\section{Uniaxially strained graphene}
\label{strain}
In a recent work\cite{ourpaper}, we explored the possibility of manipulating the indirect exchange interaction between two substitutional impurities in graphene by applying a uniaxial strain.
We found that the indirect exchange interaction between substitutional atoms separated in the armchair direction can be monotonically amplified or suppressed with uniaxial strain, while those separated in the zigzag direction displayed a more complicated, non-monotonic behaviour indicating the ability to switch off interactions between certain sublattices with strain.
Since the features of the unstrained interaction between adsorbed impurities show many differences to the substitutional case it is worth extending our study of strained graphene to include the bridge- and center-adsorbed configurations.

For uniaxial strain in the high symmetry armchair (${A}$) and zigzag (${Z}$) directions the atomic bond lengths ($R_{1/2/3}$) shown in Fig. \ref{fig:schematic} vary with the tensile strain ($\varepsilon$) applied:
\begin{equation}
 \begin{aligned}
   A\,: \, & \,  \tfrac{R_1}{R_0}  = \tfrac{R_3}{R_0}  = 1 + \tfrac{1}{4} \varepsilon - \tfrac{3}{4} \varepsilon \sigma \;, & \;\tfrac{R_2}{R_0} & = 1 + \varepsilon \; \\
   Z\,: \, & \,  \tfrac{R_1}{R_0}  = \tfrac{R_3}{R_0}  = 1 + \tfrac{3}{4} \varepsilon - \tfrac{1}{4} \varepsilon \sigma \;, & \;\tfrac{R_2}{R_0} & = 1 - \varepsilon \sigma \,,
 \end{aligned}
\label{eq:new-bonds}
\end{equation}
where $R_0 = 1.42\,\mathrm{\AA}$ is the unstrained bond length in graphene and $\sigma=0.165$ is the graphite value for Poisson's ratio, giving the level of contraction perpendicular to the direction of applied strain.
We note that we can write $R_3=R_1$ due to the symmetry of the two strain directions considered. The hopping parameters vary with bond length as
\begin{equation}
 t_i(\Delta R) = t_0 e^{- \alpha \frac{\Delta R_i}{R_0} }
\end{equation}
where $\Delta R$ is the change in the bond length, and $\alpha = 3.37$ is a constant.\cite{pereira_tight-binding_2009,PhysRevB.75.045404}
For the types of strain considered, we must therefore rewrite the Hamiltonian in Eq. \eqref{hamiltonian} and the Green function in \mbox{Eq. \eqref{eq:gab_int}} for the pristine graphene sheet, replacing the uniform hopping parameter $t$ with new parameters $t_1$ if the bond is of type $R_1$ or $R_3$ and $t_2$ if it is of type $R_2$.
This is achieved in the Green function calculation by making the substitution
\begin{equation}
 t f(\mathbf{k}) \rightarrow h(t_1, t_2,\mathbf{k}) = t_2 + 2 t_1 \, \cos k_Z \, e^{i \, k_A}
 \label{strain_f_substitution}
\end{equation}
in Eq. \eqref{eq:gab_int} and propagating it throughout the following derivations.
The analytic form of the new Green function within the SPA was calculated in Ref. \onlinecite{ourpaper} and used to determine the RKKY interaction between two substitutional impurities in strained graphene.
It is reasonably straightforward to generalise the $\Gamma_{AB}$ expressions for center-adsorbed and bridge-adsorbed impurities in a similar manner to account for the strained graphene host.
It should be noted that the applied strain will also effect the bonding between the impurity species and the graphene and may alter the magnitude of the impurity moment.\cite{Santos12, PhysRevB.84.075415, arkady_embedding_2009, machado-charry:132405}
Such effects are dependent on the exact impurity species considered and are beyond the scope of the present work, but can be expected to further influence the strain-dependent behaviour of the RKKY interaction.

In the next sections we will show the explicit strain-dependent forms of $\mathcal{M}(E, \mathbf{k}, \varepsilon)$ for center-adsorbed and bridge-adsorbed impurities which form the basis of numerical calculation of $\Gamma_{AB}$.
The strain dependence of the resultant SPA coefficients is also shown and used to explain the strain-dependent features of the indirect exchange interaction between these types of adsorbates.


\subsection{Strain effects on Center-adsorbed Impurities}
\label{strain-plaq-sec}

\begin{figure}
\includegraphics[width=0.45\textwidth]{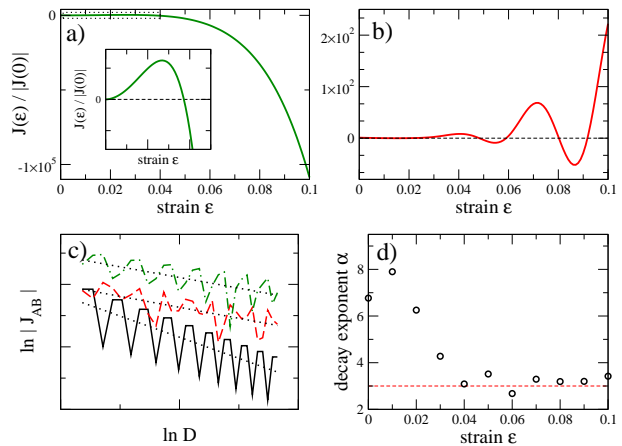}
\caption{
Numerically evaluated indirect exchange interaction $J(\varepsilon)$ between two center-adsorbed impurities fixed distances apart in the a) armchair and b) zigzag direction as uniaxial strain is applied perpendicular to the separation direction.
The results are normalised relative to the unstrained coupling $J(0)$.
The inset in panel a) shows a close-up of the region highlighted by a dotted rectangle in the main plot.
c) Log-log plots of coupling against zigzag direction separation for armchair direction strains of $\varepsilon= 0.0$ (black, solid), $0.05$ (red, dashed) and $0.1$ (green, dashed-dotted). 
The black dotted lines show linear regressions with slopes of $-6.8$, $-3.5$ and $-3.4$ respectively.
d) Decay exponent $\alpha$ as a function of strain for the cases shown in c) and additional values.
}
\label{fig:CA_strain}
\end{figure}

The strain dependent $\mathcal{M}(E, \mathbf{k}, \varepsilon)$ term for center-adsorbed impurities is found by using Eqs. \eqref{eq:mplaq} and \eqref{strain_f_substitution}. It is given by
\begin{equation}
 \mathcal{M}_C(E,\mathbf{k}, \varepsilon) = 2 E |f(\mathbf{k})|^2 +  2 \, \text{Re} \,\left[{ f^2(\mathbf{k})\, h(t_1, t_2, \mathbf{k}) \, e^{-i2k_A}}\right].
 \label{eq:mplaq_strain}
\end{equation}
Care must once more be taken that the correct sign choice for the relevant pole is made for each term in $\mathcal{M}_C(E,\mathbf{k})$ when using \mbox{Eq. \eqref{eq:mplaq_strain}} within an exact contour integral.
Using numerically evaluated Green functions, we can calculate the indirect exchange interaction for center-adsorbed impurities as a function of strain.
The top panels of Fig. \ref{fig:CA_strain} show how the coupling between center-adsorbed impurities a fixed distance apart varies as uniaxial strain is applied perpendicular to the separation direction.
The results are normalised relative to the magnitude of the coupling in the unstrained system.
We show two cases: armchair separated impurities with a zigzag strain (panel a) and zigzag separated impurities with an armchair strain (panel b).
In both cases a dramatic increase in the magnitude of the coupling is observed as the strain is increased.
It is worth noting that similar increases in the coupling magnitude, not shown here, are observed if parallel strains are 
applied.
This is in contrast to the case of substitutional impurities\cite{ourpaper}, where parallel strains are generally associated with an overall suppression of the coupling.
For the zigzag separated impurities in \mbox{Fig. \ref{fig:CA_strain}} b), we also note sizeable sign-changing oscillations, suggesting strain as a tool to manipulate the preferential spin alignment of a pair of center-adsorbed impurities.
A more subtle sign changing feature is also present for the armchair direction and highlighted in the zoomed inset of panel a, where we note the unstrained FM coupling switches to AFM initially, before returning to FM for larger values of strain.
To understand this behaviour better, we turn to the distance dependence of the coupling in strained systems.
Fig. \ref{fig:CA_strain}c) shows log-log plots of the coupling as a function of distance for zigzag-separated center-adsorbed impurities with no strain (black, solid line) and for armchair strains of $\varepsilon =0.05$ (red, dashed) and $\varepsilon=0.1$ (green, dashed-dotted).
It is clear that the slopes of the three lines are different, indicating a change in the rate of decay as 
strain is varied.
Regression fits to these curves (dotted black lines) find decay exponents of $-6.8$, $-3.5$ and $-3.4$ for the $\varepsilon = 0.0, 0.05, 0.1$ cases respectively. 
Fig. \ref{fig:CA_strain} d) plots the decay exponent, $\alpha$ (where $J \sim D^{-\alpha}$), for a number of $\varepsilon$ values.
We see that the initial unstrained asymptotic decay rate of $D^{-7}$ changes to the $D^{-3}$ rate expected for substitutional and bridge impurities (shown by a dashed red line in \mbox{Fig. \ref{fig:CA_strain}} d)) within the range $\varepsilon = 0.0 -0.05$, and it remains constant at this value for higher values of strain.
Similar transitions of the decay rate from $D^{-7}$ to $D^{-3}$ are noted for the other separation and strain directions, and explain the massive amplification of the coupling with strain noted in the top panels of \mbox{Fig. \ref{fig:CA_strain}}.

To understand the behaviour of the coupling more clearly it is worth examining the strain-dependent forms of the SPA Gamma function, and the role they play in determining the sign and decay rate of the coupling.
The strain-dependent SPA coefficients for armchair and zigzag separations are given by 
\begin{widetext}
\begin{equation}
\begin{split}
 \mathcal{A}_{C}^{ac}(E, \varepsilon) &= \tau^2\,\sqrt{\frac{2 i }{\pi}} \, \sqrt{ \frac{E}{(E^2 -t_2^2 + 4t_1^2)\sqrt{(t_2^2 - E^2)}}}\, \frac{2(E+t_2-t_1)^2 (t_2 -E) }{t_2t_1^2}\\
 \mathcal{Q}^{ac}(E, \varepsilon) &= \pm \cos^{-1}\left({ \frac{-\sqrt{t_2^2 - E^2}}{t_2}  }\right)  \\ 
\end{split}\,
\label{eq:armchair_aqs}
\end{equation}

\begin{equation}
\begin{split}
 \mathcal{A}_C^{zz}(E, \varepsilon) 
 &= \frac{\tau^2}{\sqrt{2 i \pi}}
 \sqrt{ \frac{E}{|t_2|(t_2 - E)\sqrt{(4t_1^2-(E-t_2)^2)}}}
 \frac{2(E-t_2+t_1)^2}{t_1^2} \\
 \mathcal{Q}_{C}^{zz}(E, \varepsilon) &= \pm \cos^{-1}\left({ \frac{-t_2 + E}{2t_1}  }\right)\\
\end{split}\,.
\label{eq:zigzag_aqs}
\end{equation}
\end{widetext}
It is clear that these expressions reduce to those given by Eqs. \eqref{eq:armchair_aq} and \eqref{eq:zigzag_aq} in the $\varepsilon = 0$ limit where $t_1 = t_2$.
From the discussion in Sec. \ref{plaquette} of the interaction decay rate for center-adsorbed impurities in unstrained graphene, we recall that the decay exponent $\alpha$ is determined by the order of the first non-vanishing energy derivative of $\mathcal{B} = \mathcal{A}^2$ evaluated at the Fermi energy.
From \mbox{Eq. \eqref{eq:rkkysum}}, if $\mathcal{B}^{(\ell)} \ne 0$, then $\alpha = \ell + 2$.
In the unstrained case, the first four derivatives of $\mathcal{B}$ vanish, corresponding to a decay exponent of $\alpha = 5 + 2 = 7$. The zero-th derivative, $\mathcal{B}^{(0)} = \mathcal{B}$, vanishes in both the strained and unstrained cases due to the presence of the $E$ in the numerator of $\mathcal{A}$ in Eqs. \eqref{eq:armchair_aq}, \eqref{eq:zigzag_aq}, \eqref{eq:armchair_aqs} and \eqref{eq:zigzag_aqs}. 
This is related to the vanishing density of states in graphene at the Dirac point and also occurs for substitutional and bridge-adsorbed impurities, where an $\alpha = 3$ decay is predicted for unstrained graphene.
Examining the form of $\mathcal{B}^{(\ell)}(\varepsilon)$ for center-adsorbed impurities we note that
\begin{equation}
 \mathcal{B}^{(\ell)}(E=0) \sim (t_2 - t_1)^{5-\ell} \quad \mathrm{for} \quad \ell = 1, \cdots, 5 
\end{equation}
so that the first four terms vanish in the unstrained case.
As a non-isotropic strain is applied, the quantity $t_2 - t_1$ becomes nonzero and we thus expect a decay rate of $D^{-3}$, corresponding to the $\ell = 1$ term in the series dominating, and indeed this is what we find numerically in the asymptotic case for larger values of strain. 
However, for small values of strain and small to medium separations between the impurities, the dominating term is determined by an interplay between the $(t_2 - t_1)^{5-\ell}$ term in the numerator and $D^{\ell + 2}$ term in the denominator. 
Another complicating factor is the fact that the different terms in the power series expression for the coupling may have different signs.
Thus, as strain is increased we should expect to see the decay rate decrease from $D^{-7}$ to slower decays of alternating sign before settling on $D^{-3}$ when the $\mathcal{J}_{3}$ coefficient, from \mbox{Eq. \eqref{eq:Jl}}, is large enough to dominate over those of faster decays. 
This is exactly the behaviour noted in the numerical results presented in the bottom panels of Fig. \ref{fig:CA_strain}.

We note that in addition to the sign-changing oscillations for both armchair and zigzag directions which are associated with different terms in the coupling power series dominating the interaction, another set of sign changing oscillations emerge for zigzag separations due to the strain dependence of the Fermi surface which breaks the commensurability between the oscillation period and the lattice spacing.

A similar set of oscillations, but without the sign-changing feature seen here, was noted for substitutional impurities in strained graphene.\cite{ourpaper}

%
\subsection{Strain effects on Bridge-adsorbed Impurities}

\begin{figure}
\includegraphics[width=0.48\textwidth]{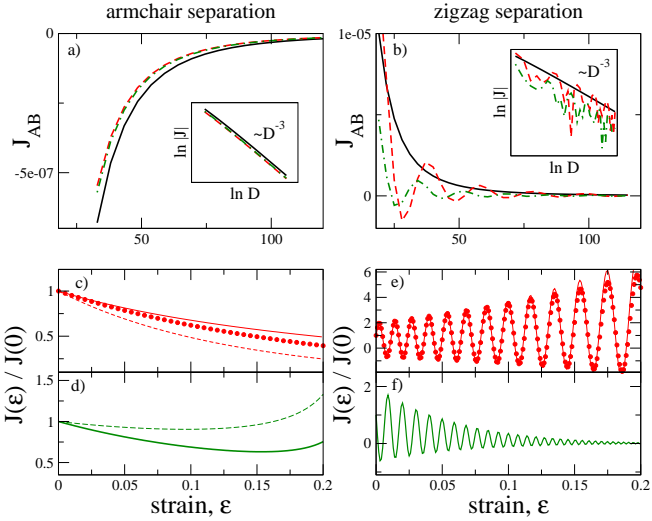}
\caption{
The effect of strain on the indirect exchange interaction between bridge-adsorbed impurities separated in the armchair (left panels) and zigzag (right panels) direction.
In all panels red (green) plots correspond to an armchair (zigzag) direction strains. a) and b) show the separation dependence of numerically calculated interactions for unstrained (black, solid) and 5\% armchair (red, dashed) or zigzag (green, dash-dotted) strains. 
The insets show log-log plots, confirming the persistence of the $D^{-3}$ decay rate.
The bottom panels show the change in the coupling, relative to the unstrained coupling, as a function of strain for fixed separations of $80\ l_A$ (c and d) and $80\ l_Z$ (e and f). 
For armchair strains (c and e) the large red dots represent numerical evaluations and the thin red lines the analytic predictions given in the text.
Only numerical evaluations are shown for zigzag strain cases (d and f). 
The dashed lines in c) and d) represent numerical evaluations for a second class of bridge atoms (see main text).
}
\label{fig:bridge_strain}
\end{figure}

The strain-dependent form of $\mathcal{M}(E, \mathbf{k}, \varepsilon)$ for bridge-adsorbed impurities is found by using Eq. \mbox{\eqref{strain_f_substitution}} to generalize \mbox{Eq. \eqref{MEk_bond}}, which yields
\begin{equation}
 \mathcal{M}_B(E, \mathbf{k}, \varepsilon) = 2E + 2  \, \text{Re} \,\left[  h(t_1, t_2, \mathbf{k}) \right]
\end{equation}
when both impurities are over an $R_2$ bond shown in Fig. \ref{fig:schematic}, and similar expressions for the other classes of bridge impurity pairs discussed in Sec. \ref{bond}.
The indirect exchange interaction between two such impurities in a strained graphene system can be calculated numerically as before, and a number of representative calculations of the coupling are presented in \mbox{Fig. \ref{fig:bridge_strain}}.
Unless otherwise stated, the bridge impurities considered sit above the $R_2$ bond. 
The top panels show the interaction between two bridge impurities as they are separated in the armchair (a) or zigzag (b) directions. 
The black curves, representing the unstrained case, are equivalent to the plots in the bottom panel of Fig. \ref{fig:bridge}, where we note that we now only consider every third separation value in the zigzag direction in order to remove the period-3 oscillations usually seen in this direction. 
The red-dashed (green dash-dotted) curve in these panels represent the interaction when armchair (zigzag) strain of strength $\varepsilon=0.05$ is applied. 
In all cases, the log-log insets in panels a) and b) reveal that, unlike for center-adsorbed impurities, strain has no 
effect on the decay rate between bond impurities, which remains at the standard $1/D^3$ rate for undoped graphene. 
For armchair separations, we note that both strains lead to a mild suppression of the coupling. 
This is in contrast to the prediction for substitutional impurities\cite{ourpaper} that parallel (armchair) strain should suppress and perpendicular (zigzag) strain amplify the coupling. 
These features are also clear when we examine the change in the coupling for impurities a fixed distance $80\ l_A$ apart as armchair (c) or zigzag (d) strain is applied.
Numerical calculations are shown by the red circles in c), and the solid line is the analytical result.
The solid green line in f) represents the numerical calculation for zigzag strains.
In both cases only suppression of the coupling is observed until high values of strain are reached.
The dashed lines in these panels represent numerical calculations performed with one of the impurities moved above an $R_1$ type bond, i.e. one of the other classes of bond impurity pairs discussed in Sec. \ref{bond}.
In this case, we note qualitatively similar behaviour for both direction strains, with only very minor suppression of the coupling for zigzag strains until amplification begins at high strain values.

For zigzag separations we note that strain induces additional sign-changing oscillations as both the separation (b) and strain (e and f) are varied. 
The oscillations with increasing separation are in addition to the existing period-3 oscillations visible in the bottom right panel of \mbox{Fig. \ref{fig:bridge}} for the unstrained case. 
For a fixed separation of $80\ l_Z$, we note that the coupling oscillates rapidly as a function of strain for both strain directions, with an overall amplification for the armchair strain (e) and suppression for the zigzag strain (f).
We note that the frequency of the oscillations increases with separation.
To better understand the behaviour for bond impurities, it is again instructive to examine the SPA form of the $\Gamma_{AB}$ term entering into the expression for the coupling.
The strained forms of the coefficients in \mbox{Eq. \eqref{eq:bond_aqs}} are 
\begin{equation}
 \begin{split}
   \mathcal{A}_{B}^{ac}(E, \varepsilon) &= \frac{- 2\tau^2(E + t_2) }{t_2} \sqrt{\frac{2 i }{\pi}}  \sqrt{ \frac{E}{(E^2 + 4t_1^2 - t_2^2)\sqrt{(t_2^2 - E^2)}}}  \\
 \mathcal{Q}^{ac}(E, \varepsilon) &= \pm \cos^{-1}\left({ \frac{-\sqrt{t_2^2 - E^2}}{t_2}  }\right)  \\ 
 \mathcal{A}_B^{zz}(E, \varepsilon) &= \frac{4 \tau^2} {\sqrt{2 i \pi}} \, \sqrt{ \frac{E}{|t_2|(t_2 - E)\sqrt{(4t_1^2 -(E - t_2)^2)}}} \\
 \mathcal{Q}_{B}^{zz}(E, \varepsilon) &= \pm \cos^{-1}\left({ \frac{-t_2 + E}{2t_1}  }\right) \,.
 \end{split}
 \label{bondstrainaqs}
\end{equation}
It is clear that the oscillations arising for zigzag direction separations are due to the strain-dependence of the Fermi wavevector $\mathcal{Q}_{B}^{zz}$ in this direction. 
This is in contrast to the armchair case, where the wavevector $\mathcal{Q}^{ac}$ is strain-independent at $E=0$.
The anisotropy of the Fermi surface under uniaxial strain has been noted previously in the literature \cite{PhysRevB.85.115432,pereira_tight-binding_2009} and is also the mechanism behind oscillations in the amplitude of the coupling noted previously for zigzag separated substitutional impurities.\cite{ourpaper}
An important difference between the bridge impurities and the substitutional case comes from the averaging out of sublattice dependent effects and the consequent possibility of either FM or AFM couplings, as seen in \mbox{Fig. \ref{fig:bridge}} and in the form of the oscillatory term in\mbox{ Eq. \eqref{bond_oscills}}.
Including the strain dependent Fermi wavevector from \mbox{Eq. \eqref{bondstrainaqs}} in \mbox{Eq. \eqref{bond_oscills}} returns the same oscillatory behaviour as calculated numerically.
In our previous study of substitutional impurities in strained systems, simple analytic expressions were derived to predict the amplification, suppression and oscillatory behaviour of $\tfrac{J(\varepsilon)}{J(0)}$ for the high symmetry directions of strain and separation.
The accuracy of these simple expressions was as a result of the simple form of the RKKY coupling expression in terms of the off-diagonal Green functions, from which the strain-dependence could be simply extracted. 
Although we have derived similar expressions for the $\Gamma_{AB}$ function in this work, the strain dependence of the coupling amplitude is complicated significantly by the fact that the denominator in \mbox{Eq. \eqref{eq:GAB}} relating the required Green functions to these $\Gamma$ terms has a non-trivial strain dependence. Focusing only on the $\Gamma_{AB}$ contribution yields analytic estimates of
\begin{align}
 \frac{J(\varepsilon)}{J(0)} & = \frac{3 t_0 t_2}{4 t_1^2 - t_2^2} \quad \mathrm{(A)}\\
 \frac{J(\varepsilon)}{J(0)} & = \frac{|t_0| \sqrt{4 t_1^2 - t_2^2}}{\sqrt{3} t_2^2} \frac{1 - 2\cos(2\mathcal{Q}(\varepsilon)D + \phi_B)}{1 - 2\cos(2\mathcal{Q}(0)D + \phi_B)}  \quad \mathrm{(Z)}\
 \label{strained_bond_ratios}
\end{align}
for armchair (A) and zigzag (Z) separations respectively.
We note that the armchair expression is identical to that for the substitutional case, and the zigzag expression varies only in the oscillatory term. 
These expressions provide a reasonable approximation for armchair direction strains, and evaluations shown by solid red lines in \mbox{Figs. \ref{fig:bridge_strain}} c) and e) match quite well with the numerical evaluations shown by the red circles. 
However the analytic expressions were found to greatly underestimate the degree of suppression noted for zigzag strains for both separation directions and are not shown in \mbox{Figs. \ref{fig:bridge_strain}} d) and f).
We emphasise however that the proper oscillatory behaviour for the coupling as a function of strain for zigzag separations is correctly predicted for both strain directions and thus \mbox{Eq. \eqref{strained_bond_ratios}} is a useful tool to predict the amount of strain required to turn off the coupling or change its sign.


\section{Conclusions}
\label{conclude}

In this work we have demonstrated that the features of the indirect exchange interaction between impurities adsorbed onto a graphene sheet differ significantly from their simpler substitutional counterparts.
In addition, the modification of these features by a simple uniaxial strain has been shown to allow an even greater degree of control over the amplitude and sign of the interaction.
The use of a composite Green function, ${\Gamma_{AB}}$, was shown to allow for a computationally efficient calculation of this interaction in both strained and unstrained cases.

Specifically, we have shown that the RKKY interaction between adsorbed magnetic moments in graphene depends on the exact adsorption configuration of the impurities, decaying with separation $D$ as $D^{-7}$ for center-adsorbed impurities and $D^{-3}$ for bridge-adsorbed impurities, with bridge-adsorbed impurities also displaying a sign changing behaviour as a function of separation in the zigzag direction.
Using our prescription, the decay, along with other features of the interaction, may be derived in a mathematically transparent fashion.

We have also shown, analytically and numerically, that mechanical strain modifies the RKKY interaction.
Symmetry breaking of the hexagonal lattice by uniaxial strain leads to a significantly slower decay rate between center-adsorbed impurities ($D^{-3}$), which introduces the possibility of dramatically amplifying the interaction between them.
Bridge-adsorbed impurities separated along certain directions alternate between ferromagnetic and anti-ferromagnetic coupling as a function of separation and applied uniaxial strain introduces further sign changing features.
Such strain dependent behaviour suggests the intriguing possibility of selectively tuning the coupling between moments.

Since a whole range of physical features, such as magnetotransport and overall magnetic moment formation, are predicated upon the magnetic coupling, it is hoped that this work will help clarify some the discrepancies in the literature.
Experiments to date searching for magnetism in disordered graphene seem to suggest paramagnetic, non-interacting moments\cite{sepioni_limits_2010}.
Signatures of indirect exchange interactions between such moments in graphene are very difficult to detect due to their short ranged nature, particularly if they adopt certain adsorption configurations, as we have demonstrated here.
Amplification of these couplings using strain may provide a path to their detection in future experiments.
The strain dependent features predicted in this work may also find applications in carbon-based spintronics, where the ability to selectively tune the coupling between transition-metal adsorbates using strain introduces an additional degree of freedom in the characterisation of graphene spintronic devices.


\begin{acknowledgments}
The authors acknowledge financial support received from the Programme for Research in Third-Level Institutions PRTLI5 Ireland, the Irish Research Council for Science, Engineering and Technology under the EMBARK initiative and from Science Foundation Ireland under Grant No. SFI 11/RFP.1/MTR/3083. 
The Center for Nanostructured Graphene (CNG) is sponsored by the Danish National Research Foundation, Project No. DNRF58.
\end{acknowledgments}

\end{document}